\documentclass[sigconf,nonacm]{acmart}

\renewcommand\footnotetextcopyrightpermission[1]{}

\usepackage{booktabs}
\usepackage{listings}
\usepackage{etoolbox}
\usepackage{xspace}

\usepackage[htt]{hyphenat}

\BeforeBeginEnvironment{lstlisting}{\par\noindent}

\begin{document}

\title{Intent-Based Cryptographic API Design\\for Cryptographic Agility}

\author{Navaneeth Rameshan}
\email{vme@zurich.ibm.com}
\affiliation{\institution{IBM Research Europe}\city{Zurich}\country{Switzerland}}

\author{Gregoire Messmer}
\email{Gregoire.Messmer@ibm.com}
\orcid{0000-0000-0000-0000}
\affiliation{\institution{IBM Research Europe}\city{Zurich}\country{Switzerland}}

\renewcommand{\shortauthors}{Rameshan and Messmer}

\begin{abstract}

As organizations move toward post-quantum cryptography, they face the major challenge of updating cryptographic algorithms across large, complex software portfolios. However, most cryptographic APIs in use today were designed around specific algorithms. These APIs expect explicit use of specific algorithms, provide little or no support for policy-based algorithm selection, and offer no straightforward way to migrate existing keys to newer algorithms. This makes the transition to post-quantum cryptography challenging. The companion assessment framework~\cite{paper1-framework} identifies the barriers for cryptographic agility and explains why algorithm transition is largely a software engineering problem.

To address the limitations of current cryptographic APIs, we identify the principles necessary to design a cryptographically agile API. The design principles are derived from five fundamental architectural characteristics (Abstraction, Stability, Temporal Flexibility, Separation, and Extensibility). We also show how the design principles can be implemented using several examples of Protocol Buffers API design patterns. In particular, we present an intent vocabulary that is based on scopes which allows for decoupling key creation from algorithm identities. It also supports transparent substitutions of algorithms in the applicable scope. Cryptographic governance is enabled by an abstract policy API that does not prescribe the policy format. Keys are represented by stable identifiers and support key evolution operations (rotation, transformation, migration), facilitating migration between algorithms and providers while tracking both the original key identity and its evolution history. With this approach, updating cryptography becomes an operational process without the need to rewrite application code.

\end{abstract}

\maketitle


\section{Introduction}

\label{sec:introduction}


The National Institute of Standards and Technology (NIST) has standardized several post-quantum cryptographic algorithms, including FIPS 203 (ML-KEM)~\cite{fips203}, FIPS 204 (ML-DSA)~\cite{fips204}, and FIPS 205 (SLH-DSA)~\cite{fips205}. This decision represents the most significant cryptographic transition ever undertaken in deployed systems. To facilitate this shift, NIST has established specific timelines: 112-bit classical algorithms will be phased out by 2030 and completely prohibited by 2035~\cite{nist-sp800-131a,nist-sp1800-38}. 

The primary challenge in this transition arises from the system architecture rather than the cryptographic algorithms themselves. Cryptographic algorithms are typically embedded directly in the application's source code~\cite {lazar2014,acar2017}, and updating them often becomes a large software engineering project that requires modifications across multiple source files. In the past, transitions such as the move from SHA-1 to SHA-256 or from DES to AES, have shown that the greatest difficulty lies not in selecting new algorithms but in the effort required to locate, update, test, and redeploy every affected application. The transition to post-quantum cryptography is much more complex because the new algorithms also introduce features such as increased key sizes, larger signature outputs, and revised parameter structures. Consequently, these changes may also necessitate adjustments to how data is processed, stored, or transmitted throughout the system.

\subsection{The Agility Gap in Deployed Systems}

The companion paper~\cite{paper1-framework} introduces a comprehensive framework to assess the agility of cryptographic APIs and divides it into seven distinct dimensions. This framework is then used to conduct a detailed analysis of six widely used cryptography APIs: PKCS\#11, OpenSSL 3.0, Java Cryptography Architecture (JCA), Google Tink, AWS Key Management Service (KMS), and HashiCorp Vault Transit. The evaluation examines the practical capabilities of each system and identifies three recurrent deficiencies across all the systems analyzed.

\begin{enumerate}

\item \textbf{Absent intent-based key creation.} None of the deployed systems evaluated supports the ability to specify cryptographic intent (such as a requirement for authenticated encryption, for example). In all six systems, users must provide algorithm-specific identifiers when creating cryptographic keys. As a result, migrating to a new algorithm requires code changes at each deployment site.

\item \textbf{Absent cryptographic governance.} Only AWS Key Management Service (KMS) and HashiCorp Vault implement role-based access control, and neither system provides cryptographic governance. Role-based access control and cryptographic governance are distinct: the former determines who can access a cryptographic key, while the latter enforces policies on which algorithms and configurations are permitted, which is essential for meeting security and compliance requirements.

\item \textbf{Absent algorithm transformation.} None of the application programming interfaces (APIs) evaluated in this study exposes an explicit operation that allows the algorithm used by a cryptographic key to be updated while preserving the same key identity, which is essential for achieving cryptographic agility.

\end{enumerate}

These gaps are not simply accidental mistakes in how the systems were built. They come from deeper issues in the original API designs. Fixing them will require improving the design itself to address these core problems.

\subsection{Paper Organization}

Section~\ref{sec:framework-recap} provides a concise recap of the assessment framework. Section~\ref{sec:principles} derives the thirteen design principles from five foundational properties for an agile cryptographic API. Section~\ref{sec:api} demonstrates their realization through Protocol Buffers API patterns. Section~\ref{sec:evaluation} evaluates the API against the assessment framework and Section~\ref{sec:migration} presents an end-to-end migration scenario. Finally, Section~\ref{sec:related} reviews the related work and Section~\ref{sec:conclusion} concludes.


\section{Assessment Framework Recap}

\label{sec:framework-recap}


This section provides an overview of the assessment framework introduced in the companion paper~\cite{paper1-framework}. The framework decomposes cryptographic agility into seven orthogonal dimensions, each evaluated on a scale from 0 to 4. Comprehensive definitions and constraints, as well as the evaluation of widely used cryptographic APIs, are presented in the companion paper.

\subsection{Framework Components}

\noindent\textbf{Tier 1: Decoupling Assessment.} This tier examines the degree to which the cryptographic API enforces specific cryptographic choices. It is composed of three primary dimensions: \textbf{C1} (operation coupling), which assesses the degree to which cryptographic algorithms or properties are expected to be explicitly specified during cryptography operations; \textbf{C2} (creation coupling), which evaluates the degree to which key generation expects particular algorithms; and \textbf{C3} (provider coupling), which determines the expectation and dependency on where the cryptography is executed. In addition, \textbf{C4} (decoupling mechanism) evaluates how these dependencies are externalized across all three coupling dimensions. Finally, \textbf{C5} (authority) measures the level of support of the entity that exercises control over the decoupling mechanism.

\noindent\textbf{Tier 2: Enabler Assessment.} This tier measures the capabilities of the API to change algorithms and migrate keys between different execution environments. Specifically, the two dimensions are: \textbf{E1} (algorithm migration), which evaluates the capability to version, transform, and replace cryptographic algorithms; and \textbf{E2} (provider migration), which assesses the ability to transfer cryptographic keys between providers.

\subsection{Key Levels Referenced in This Paper}

The design principles and API patterns discussed in this paper correspond to specific components of the system:

\begin{itemize}

\item \textbf{C1.2} (key-driven dispatch): operations are selected based on the key itself, not by passing algorithm choices each time.

\item \textbf{C1.3} (scope-consistent operations): the same set of parameters works across all algorithms within a given scope.

\item \textbf{C2.2} (intent-based creation): instead of naming algorithms directly, applications describe what they want to achieve using a shared vocabulary.

\item \textbf{C3.2+} (uniform API abstraction): a single interface hides differences between cryptographic providers.

\item \textbf{C4.3} (policy engine): rules define cryptographic requirements, with checks to ensure they make sense semantically.

\item \textbf{C5.3+} (scoped governance): different parts of an organization can define and manage their own cryptographic rules and responsibilities.

\item \textbf{E1.3} (explicit algorithm transformation): a dedicated operation migrates a key to a different algorithm.

\end{itemize}

\subsection{Architectural Vocabulary}

The framework establishes four key abstractions that together constitute a binding chain: a \textit{scope}, which represents a named intent category grouping algorithms with compatible operational interfaces; a \textit{template}, defined as an immutable, named algorithm configuration with specified lifecycle states; a \textit{policy}, which serves as a governance instrument for controlling cryptographic posture; and a \textit{provider}, which functions as the execution backend. Applications articulate their intent through a scope; policies then determine which templates are permissible, and the system subsequently dispatches operations to an appropriate provider. This structure enables agility, as each link in the chain can be modified independently without disrupting the overall system.

\subsection{The Three Gaps}

Evaluation of six systems~\cite{paper1-framework} reveals three key gaps: (1) no system achieves C2.2 (intent-based creation); (2) no system achieves C5.3+ (cryptographic governance); and (3) no system achieves E1.3 (algorithm transformation). The principles and patterns presented in the following sections are intended to address these shortcomings.


\section{API Design Principles for Cryptographic Agility}

\label{sec:principles}


The assessment framework describes a system's position along each dimension of cryptographic agility. This section addresses the related question of which API design provisions are necessary to progress along those dimensions. We derive thirteen principles by analyzing the functional requirements imposed by each framework component on API surfaces. We identify five foundational properties that underlie these requirements, and capture the resulting obligations as design principles.

\subsection{Foundational Requirements for Cryptographic Agility}

We identify five core properties that jointly form the foundation of agility for a cryptographic API. These properties define the main requirements that any agile cryptographic API must satisfy.

\textbf{Abstraction} is the separation of cryptographic details from the developer-facing API. Applications should be able to express their intent to use cryptography without being tied to specific algorithms, key representations, cryptographic properties, or execution environments. Abstraction allows cryptographic mechanisms to change without affecting application logic or requiring application developers to understand low-level cryptographic details.

\textbf{Stability} means that applications' use of the cryptographic API remains consistent even when algorithms, providers, or policies change. This helps ensure that updates to cryptography do not require changes in applications that depend on it. API Stability is mainly about preserving backward compatibility and reliable behavior for existing applications.

\textbf{Temporal Flexibility} means cryptographic decisions can be made or changed at any stage of an application's lifecycle. An agile cryptographic API should support setting and enforcing cryptographic behavior during the design, deployment, and operation phases. This makes it possible for cryptographic choices to adapt as requirements change over time.

\textbf{Separation} enables distributing cryptographic responsibilities among distinct domains of authority.  Application development, security governance, and operational management in an organization are decentralized responsibilities and should remain independent. This separation allows each authority to control its own domain. The API should provide clear interfaces for different stakeholders to fulfill their responsibilities.

\textbf{Extensibility} describes a system's ability to support new cryptographic capabilities and requirements without major architectural change. When new algorithms, standards, providers, or deployment models become available, the API should be able to incorporate them while preserving existing abstractions and interfaces. This helps the system evolve over time. Extensibility is primarily concerned with forward compatibility and the ability of the system to grow without redesign.

While stability and extensibility are closely related, they address different concerns. Stability ensures that existing applications continue to function without modification when underlying cryptographic components change. Extensibility guarantees that new cryptographic capabilities can be introduced without requiring changes to the core architecture. An agile cryptographic API must balance both these properties: it must remain stable for existing consumers while allowing future evolution.

\subsection{ Design Principles}

The five properties describe the essential qualities of an agile API. The principles explain the actions or requirements needed to achieve these qualities. Each principle is based on at least one of the core properties. We present thirteen principles. Each one is organized according to the property it supports. We also classify these principles as either core or supportive, depending on their role in achieving agility.

\noindent\textbf{Core Principles} are necessary conditions for cryptographic agility. Violating any core principle prevents safe substitution of algorithms or providers. In contrast, \textbf{Supportive Principles} improve usability, governance, and long-term maintainability but are not strict prerequisites for agility.

\subsubsection{Core Principles}

\noindent\textbf{P1: Operation Independence} (\textit{Abstraction}). \textit{The API must define cryptographic operations independently of the algorithms and their properties.} The API must expose all cryptographic operations such as encryption, signing, key derivation, etc., without requiring algorithm-specific parameters or algorithm identifiers. Any parameters exposed by the API must be semantically shared across all algorithms that implement the same operation. This principle enables C1.2+ decoupling by ensuring that operations remain stable under algorithm substitution.

\noindent\textbf{P2: Temporal Decoupling} (\textit{Temporal Flexibility}). \textit{The API must allow cryptographic decisions to be made at different lifecycle stages of the application that consumes it.} The API must allow decisions to be made at design time (operation intent specification), deployment time (algorithm and provider selection), and runtime (key selection). Changes at one stage must not require changes to earlier stages. This principle enables C4.1+ and C4.3+ decoupling.

\noindent\textbf{P3: Intent-Based Specification} (\textit{Abstraction}). \textit{The API should expose the ability to specify the cryptographic intent of a user instead of being bound to an exact choice in the algorithms used.} The API must allow callers to specify their intent, required security properties and constraints, such as compliance requirements or quantum resistance, without requiring explicit algorithm selection. The API must be responsible for resolving valid implementations that satisfy the expressed intent. This principle enables C2.2+ creation decoupling by separating intent specification from implementation binding.

\noindent\textbf{P4: Operational Parameter Consistency} (\textit{Stability}). \textit{The API must ensure that all cryptographic operations expose the same operational parameters for a given intent.} The API must guarantee that substituting one algorithm for another, during a cryptographic operation, within the same intent, exposes the same set of input/output parameters. If algorithms require incompatible operational parameters, the API must not classify them under the same intent. This principle ensures safe substitution within a given intent and preserves stable API behavior across algorithm changes.

\noindent\textbf{P5: Key Abstraction} (\textit{Stability}). \textit{The API must expose cryptographic keys via stable identifiers and be independent of algorithm and implementation details.} Keys should be represented by identifiers and must retain the same identity across algorithm changes, provider changes, and key lifecycle events. Identifiers must represent logical key purposes rather than implementation details. This principle enables C1.2, E1.1+, and E2.1+ by ensuring stable key references across all changes.

\noindent\textbf{P6: Key Evolution} (\textit{Temporal Flexibility}). \textit{The API must support key evolution operations without disrupting existing operations.} The API must provide mechanisms for key rotation (new key material with the same algorithm), algorithm transformation (new key material with a new algorithm or cryptographic property), and provider migration (migrating key to a different provider) while preserving continuity of cryptographic operations and compatibility with existing data. This principle enables E1.1+ and E2.1+ by ensuring lifecycle evolution is supported at the API level.

\noindent\textbf{P7: Separation of Roles}\textit{(Separation)}: \textit{The API must permit separation of responsibility for cryptographic use, governance, and operational management}. The API must define separate interfaces or roles to invoke cryptographic operations, set policy, and configure an execution environment. Changes in governance or configuration should not necessitate a modification to how the application uses the API interface. This principle supports C5.3+ as it enforces role separation at the API boundary.

\noindent\textbf{P8: Algorithm--Implementation Separation} (\textit{Separation}) [\textit{actionable at C3.2+}]. \textit{The API must decouple algorithm selection from the execution environment in which the cryptography runs.} The API must allow algorithms to be selected independently of where or how they are executed, and execution environments must not constrain the set of allowed algorithms. This principle enables C3.2+ by separating algorithm choice from execution context within the API.

\noindent\textbf{P9: Implementation Portability} (\textit{Abstraction}) [\textit{actionable at C3.2+, E2.1+}]. \textit{The API must allow substitution of cryptographic implementation without changes to the application code.} The API must present a uniform interface regardless of the execution backend, and key material must be transferable between providers where security models permit. This principle enables C3.2+ and E2.1+ by ensuring portability across implementations through a stable API contract.

\noindent\textbf{P10: Policy Independence} (\textit{Extensibility}) [\textit{actionable at C4.3+}]. \textit{The policy engine implementation for cryptographic governance must be pluggable.} The API should reference policies by name (e.g., \texttt{policy: "production-fips"}) without mandating a specific syntax, language, or evaluation engine. The policy abstraction requires a well-defined interface for validating requests and enforcing allow/deny lists. This principle enables C4.3+ by ensuring policy evolution is external to application API contracts.

\subsubsection{Supportive Principles}

\noindent\textbf{P11: Language Independence} (\textit{Abstraction}). \textit{The API must be defined independently of a specific programming language for widespread applicability.} The API must provide a language-neutral specification of cryptographic operations to ensure usability across different programming languages. This principle supports C1–C2 by ensuring consistent API behavior across heterogeneous environments.

\noindent\textbf{P12: Property Taxonomy} (\textit{Extensibility}). \textit{The API must expose cryptographic algorithms and providers through a structured and consistent property model.} The API must define properties in a way that supports consistent reasoning about suitability for different security and operational requirements. This principle supports C4 by enabling structured and extensible algorithm selection through the API.

\noindent\textbf{P13: Discoverable Extensibility} (\textit{Extensibility}). \textit{The API must allow new cryptographic algorithms and providers to be added and discovered without changing the existing API definitions.} The API must support controlled introduction, visibility, and retirement of cryptographic capabilities while maintaining backward compatibility. This principle enables C4.3+ and ensures that the API remains stable even when cryptographic capabilities change.

\subsection{Principle to Component Mapping}

The core principles cover all seven dimensions of the assessment framework in the companion paper. P1–P4 mainly reduce algorithm coupling in operations (C1) and key creation (C2). P5–P6 focus on key management and how algorithms can be migrated over time (E1). P7 and P10 deal with governance (C5) and the policy layer (C4.3+). P8–P9 address coupling to providers (C3) and support moving between providers (E2). The remaining three supportive principles (P11–P13) don’t define core requirements, but they make the system easier to use, extend, and roll out in practice.


\section{API Design Patterns: Principle Realization}

\label{sec:api}


This section presents a concrete API design that uses Protocol Buffers patterns. For each component in the framework of the companion paper, we present the coupling problem, the API pattern that addresses it, and the design principles realized. The patterns generalize to other interface definition languages and provide reproducible guidance for practitioners.

\noindent\textbf{Architectural Vocabulary.} The API relies on four architectural abstractions that collectively form a binding chain through which application intent is resolved to cryptographic execution.

A \textit{scope} is a named intent category that groups algorithms sharing compatible operational interfaces. Each scope corresponds to a cryptographic primitive (e.g., authenticated encryption, digital signature) and guarantees that all algorithms within the scope accept the same operational parameter as input. Scopes are the mechanism by which applications express \textit{what} security property they need without specifying \textit{which} algorithm provides it.

A \textit{template} is an immutable, named configuration that fully specifies a concrete algorithm and all its parameters---key size, mode of operation, hash function, padding scheme etc.. Templates are registered in a catalog with lifecycle states (experimental, active, acceptable, deprecated, and forbidden), enabling controlled transitions between algorithms. Each template also declares the scopes it satisfies.

A \textit{policy} is a governance instrument that controls the cryptographic posture of the system. Policies define which templates are permitted and forbidden, constrain key lifecycle parameters, restrict which providers may be used, and determine which template satisfies a given scope and set of constraints, when scope-based key creation is used. The policy engine is an abstract interface. The API makes no assumptions about the policy schema or the policy evaluation logic.

A \textit{provider} is an execution backend that performs the actual cryptographic computation. For example, a provider could be a software library, an HSM, a cloud KMS, or a remote cryptographic service.

These four abstractions compose into a resolution chain: \textit{scope} $\to$ \textit{policy} $\to$ \textit{template} $\to$ \textit{provider}. 

\subsection{The Binding Chain}

\label{sec:binding-chain}

Application specify their intent for cryptography use through a \texttt{scope}. The policy then selects a concrete \texttt{template} and a \texttt{provider} that fulfills the scope. The selected \texttt{template} is then dispatched to a \texttt{provider} to execute the cryptography. The binding occurs at two points: at \textit{key creation time}, when intent is resolved to a concrete template and provider. Then, at \textit{operation time}, a cryptographic operation is dispatched to a provider for execution. 

In particular, at key creation, an implementation proceeds in the following manner: (1)~the application requests a key with a scope and optional properties; (2)~the policy engine selects a permitted template; (3)~the template registry confirms the template; (4)~the provider registry identifies a capable provider that supports the template; (5)~the key is generated using the provider and the metadata is stored. Then, when the application wants to execute a cryptographic operation, it provides the key name and data. The template and the provider is from the stored key metadata and the cryptographic operation is executed with the provider.

\subsection{Key Creation Decoupling (C2)}

\label{sec:api-creation}

The framework identifies creation coupling as the most pervasive gap: every deployed system requires algorithm-specific identifiers at key creation. This subsection traces the reasoning from cryptographic primitives to an intent vocabulary based on operational parameters. It realizes principles P3 (Intent-Based Specification), P4 (Operational Parameter Consistency), P5 (Key Abstraction), and P12 (Property Taxonomy).

\subsubsection{Primitives as Security Contracts.}

A cryptographic primitive defines a contract of security properties that every conforming algorithm must satisfy. The AEAD primitive guarantees confidentiality, integrity, and authenticity. Digital signatures provide integrity, authenticity, and non-repudiation. This invariance makes the primitive a meaningful unit of abstraction: selecting a primitive is selecting a set of security guarantees.

Primitives, therefore, seem to be a natural vocabulary for intent-based key creation. An application can declare its intent for authenticated encryption without naming an algorithm using the AEAD primitive. This observation aligns with Tink's primitive-based API and CycloneDX's parallel taxonomy of primitives.

\subsubsection{The Operational Interface Problem.}

Primitives group algorithms by security properties. Algorithms within the same primitive, however, may require \textit{different inputs} from the application. Table~\ref{tab:param-divergence} illustrates cases of incompatible operational interfaces within the AEAD and digital signature primitives.

\begin{table}[t]

\caption{Operational parameter divergence within cryptographic primitives.}

\label{tab:param-divergence}

\centering

\footnotesize

\begin{tabular}{@{}lll@{}}

\toprule

\textbf{Algorithm} & \textbf{Primitive} & \textbf{Input} \\

\midrule

Ed25519 & Digital Signature & message \\

ML-DSA with Context & Digital Signature & message, context \\

AES-GCM & AEAD & message, AAD (optional) \\

AES-SIV & AEAD & message, AAD (recommended) \\

\bottomrule

\end{tabular}

\end{table}

Cryptographic agility requires more than substituting algorithms that implement the same primitive category. Algorithms within a single primitive may impose different operational contracts on the caller, even when they provide equivalent high-level functionality. For example, within the digital signature primitive, Ed25519 signs raw messages directly, whereas prehashed variants such as Ed25519ph require the caller to provide a precomputed digest. Other schemes, such as ML-DSA with context binding, additionally require an explicit context parameter during signing and verification. Although these algorithms all implement digital signatures, they are not operationally interchangeable: an application designed to sign raw messages without contextual binding cannot transparently substitute an algorithm that requires prehashed input or mandatory context data.

Similar distinctions arise in other primitives. For example, some authenticated encryption APIs expect the caller to supply a nonce for each operation, whereas others manage nonce generation internally and expose no nonce parameter at all. These differences do not alter the primitive category, but they do alter the operational interface presented to applications during cryptographic operation. This means that algorithm substitution requires not only primitive compatibility but also operational interface compatibility. The substituted algorithms must impose the same responsibilities and input expectations on the caller. This requirement motivates subdividing primitives into scopes, where each scope defines a stable invocation contract that remains invariant under algorithm substitution.

\subsubsection{From Primitives to Scopes.}

We saw that the input parameters for algorithms can vary within a primitive. Therefore, we subdivide primitives into \textit{scopes} in a manner that groups algorithms within a scope to have the same set of inputs. A scope defines the caller-visible invocation contract of a primitive: which inputs must be supplied, which are optional, and which are forbidden. A scope, therefore, defines the complete contract between the application and the cryptographic API. Algorithms are characterized independently by their intrinsic cryptographic properties—such as determinism, nonce-misuse resistance, or prehash support. These are expressed through selection filters during algorithm resolution. A policy engine may therefore substitute algorithms only when both the primitive type and the operational scope are compatible. This separation preserves cryptographic agility while preventing unsafe substitutions between algorithms that have incompatible caller behavior.

We classify cryptographic parameters so that they fall into at least one of three categories:
Algorithm parameters, System-generated values, and Operational parameters. They are described further in \ref{sec:api-operation}. Operational parameters are supplied per cryptographic operation and can only be provided by the application's user. Examples include additional authenticated data, External IVs, domain-separation context, XTS tweaks, KDF inputs, etc. These encode application-level meaning that the system cannot generate.

{\emergencystretch=1em

Scope boundaries are defined around this category (operational parameters) because only these are visible to the application at operation time. Two algorithms that require different operational parameters present different application contracts and belong in different scopes. Two algorithms that require the same operational parameters are substitutable regardless of how they differ in algorithm parameters or system-generated values. Table~\ref{tab:sig-scopes} shows the scopes for digital signatures and the operational parameters that the application is expected to provide.\par} 

\begin{table}[t]

\caption{Signature scopes derived from operational parameter divergence.}

\label{tab:sig-scopes}

\centering

\footnotesize

\begin{tabular}{@{}lll@{}}

\toprule

\textbf{Scope} & \textbf{Operational Parameters} & \textbf{Example Algorithms} \\

\midrule

\textsc{sig\_standard} & message & ECDSA, RSA-PSS, Ed25519 \\

\textsc{sig\_context} & message, context & Ed25519ctx, ML-DSA \\

\textsc{sig\_prehashed} & digest & Ed25519ph, ECDSA \\

\bottomrule

\end{tabular}

\end{table}

Within each scope, the policy can substitute any algorithm for another without requiring changes to the application code. This guarantee transforms scopes from a naming convention into a \textit{substitutability contract}.
The API defines scopes as enumerations within a primitive and not as a single flat type. Every cryptographic primitive contains its own scope enum. A \texttt{ScopeSpecification} message composes them through a \texttt{oneof} so that only a single scope can be specified at any time for a key:

\begin{lstlisting}

enum SignatureScope {
 STANDARD = 1; // ECDSA, RSA-PSS
 WITH_CONTEXT = 2; // Ed25519ctx, ML-DSA
 PREHASHED = 3; // Ed25519ph
}
enum AeadScope {
 WITH_AAD = 1; // AES-GCM, AES_SIV, ChaCha20Poly1305
 STREAMING = 2; // Chunked AEAD
 WITH_EXTERNAL_IV = 3;
}
enum SymmetricCipherScope {
 BLOCK = 1; // AES-CBC, 3DES
 STREAM = 2; // AES-CTR, OFB
}
enum DiskEncryptionScope {
 STANDARD = 1; // AES-XTS (tweak)
}
enum KdfScope {
 EXTRACT_EXPAND = 1; // HKDF-style
 PASSWORD = 2; // PBKDF2, Argon2
 AGREEMENT = 3; // Key agreement derived
 COUNTER = 4; // SP800-108
 TLS = 5; GOST = 6; 
 VENDOR = 7;
}
// Many more scopes, such as:
// MAC, KEM, KeyAgreement, Hash, KeyWrapping, etc.

message ScopeSpecification {
 oneof scope_spec { // Exactly one primitive
  SignatureScopeSpec signature = 1;
  AeadScopeSpec aead = 2;
  MacScopeSpec mac = 3;
  KemScopeSpec kem = 4;
 }
}

\end{lstlisting}

Per-primitive typing provides two advantages over a flat enumeration. First, each primitive's scope values occupy an independent namespace, so \texttt{STANDARD\,=\,1} carries distinct, primitive-specific semantics enforced by the type system at compile time. Second, each \texttt{ScopeSpec} message carries typed security properties specific to its primitive---for instance, \texttt{AeadScopeSpec} exposes a \texttt{nonce\_misuse\_resistant} flag while \texttt{SignatureScopeSpec} exposes a \texttt{deterministic} flag---enabling callers to express constraints that are meaningful only within that primitive without resorting to untyped string maps.

\subsubsection{Template and Property-based Filtering.}

Each of the scopes in our architecture have \emph{concrete templates} which completely define all the algorithm parameters and are immutable. As an example, the template \texttt{aes-256-gcm-96-128} defines an algorithm family, a key-size, a mode, an IV-length, and a tag-length. Our templates use CycloneDX CBOM naming conventions for their names, and are assigned lifecycle state (\textsc{experimental} $\rightarrow$ \textsc{active} $\rightarrow$ \textsc{acceptable} $\rightarrow$ \textsc{deprecated} $\rightarrow$ \textsc{forbidden}).

\emph{Properties} are a set of named, typed characteristic descriptors that can be used to select a subset of templates or providers, or are purely informational and not used for filtering. We identify four categories (P12) of properties:

\noindent\textbf{(i)} \emph{Output-characteristic properties} are those which describe observable behaviors of an algorithm that can vary over different templates within the same scope (e.g. \texttt{deterministic}, \texttt{nonce\_misuse\_resistant}).

\noindent\textbf{(ii)} \emph{Universal security properties} are those which can be used to select a subset of templates based upon universal aspects related to security (e.g. \texttt{fips\_approved}, \texttt{quantum\_safe}, \texttt{security\_strength\_bits}).

\noindent\textbf{(iii)} \emph{Informational properties} are those which are invariant with respect to a particular scope and include information useful for auditing purposes and/or generating a CBOM (confidentiality, integrity and authenticity).

\noindent\textbf{(iv)} \emph{Implementation properties} describe the characteristics of a provider's implementation (the level at which the provider complies with FIPS 140-3, whether or not the implementation was formally verified, etc.) and reside on the provider's record, rather than being included in templates. Together, the scope and properties represent the complete intent to perform a task without referencing any specific algorithm.

\subsubsection{The Key Creation Interface.}

\begin{lstlisting}

message CreateKeyRequest {
  string name = 1;    // Human-readable identifier
  string policy = 2;  // e.g., "production-fips"
  string provider_id = 3;
  map<string, string> provider_configuration = 4;
  ScopeSpecification scope_spec = 5;
  string template_id = 6;  // Explicit and optional
  ProviderRequirements provider_requirements = 7;
}

\end{lstlisting}

The application provides a scope and optional security properties; the policy engine selects a permitted template satisfying those constraints. Optionally, the application can name an exact template and the policy validates permission and uses the specified template.

A critical structural decision is that the security properties reside \textit{inside} each primitive's \texttt{ScopeSpec} message rather than in a separate top-level map. For example, the \texttt{AeadScopeSpec} has both a \texttt{nonce\_misuse\_resistant} field (an output-characteristic property scoped by the AeadScopeSpec) and a security field containing a \texttt{UniversalSecurityProperties} with fields like \texttt{quantum\_safe}, \texttt{fips\_approved}, and \texttt{security\_strength\_bits}. By placing the properties under the scope specification, we ensure that properties are structurally bound to the scope they constrain. Additionally, the use of typed property fields provides a compile-time check --- for example, a caller cannot ask for \texttt{nonce\_misuse\_resistant} on a \texttt{SIGNATURE\_STANDARD} scope, because it exists only on AeadScopeSpec. Thus, filtering templates can be expressed through these typed fields: a request asking for scope \texttt{SIGNATURE\_STANDARD} with \texttt{security.quantum\_safe = true} limits the choice to post-quantum signature algorithms without explicitly referencing any specific algorithm.

The  \texttt{provider\_requirements} field allows us to separate concerns. Compliance requirements ("must use FIPS 140-3 Level 3") can be separated from concerns regarding algorithm requirements. Similarly, the optional field \texttt{provider\_id} enables applications to directly address providers when necessary. Finally, the field  \texttt{provider\_configuration} is intended to allow applications to provide deployment-specific configuration parameters (e.g. HSM partition id, slot number). The field name is the key identifier that will remain valid after rotation of algorithms, migration of providers and changes to policies.

\subsection{Operation Decoupling (C1)}

\label{sec:api-operation}

The API exposes only a key name, data and scope-specific operational parameters during a cryptographic operation. This represents the fundamental concepts behind P1 (Independence of Operation), P4 (Consistency of Operational Parameters), and P5 (Abstraction of Key).

\subsubsection{Uniform Key Driven Dispatch with Consistent Scope-Specific Parameters.}

All cryptographic operations that require a key and data follow a uniform structure defined as follows: key name, user data,  and scope-specific parameters. The key serves as a stable identifier and encapsulates its metadata regarding algorithm-specific properties via templates. The system determines the template, algorithm parameter, and provider for the key identifier. Each scope defines its own set of scope-specific parameters via a single\texttt{oneof scope\_params} field. The name of this oneof field is fixed throughout all operation types, and the actual contents of the message within it are specific to each scope. This uniformity makes it simple and consistent for developers to work with. The same structural convention is followed for any new operation type.

\begin{lstlisting}

message SignRequest {
  string key_name = 1;
  bytes input = 2;
  oneof scope_params {
    NoParams no_context = 3;
    SignatureDomainContext domain_context = 4;
    VendorSignatureContext vendor_context = 5;
  }
  map<string, string> user_context = 15;
}

message EncryptRequest {
  string key_name = 1;
  bytes plaintext = 2;
  oneof scope_params {
    NoParams no_params = 3;
    AeadEncryptParams aead_params = 4;
    XtsEncryptParams  xts_params  = 5;
    AsymmetricEncryptParams asymmetric_params = 6;
    VendorEncryptionParams vendor_params = 7;
  }
  map<string, string> user_context = 15;
}

message AeadEncryptParams {
  bytes additional_authenticated_data = 1;
  optional uint64 plaintext_length = 2;
}

message SignatureDomainContext {
  bytes context = 1;  // Domain-separation string
}
\end{lstlisting}

The \texttt{oneof scope\_params} structure is defined by the scope, and not by the algorithm family. Algorithms in the standard signature scope require the field \texttt{NoParams}. It represents an empty message that explicitly declares that no additional parameters are needed. This \texttt{NoParams} pattern is required to be specified explicitly and is a deliberate design choice. In Protocol Buffers, an unset \texttt{oneof} is indistinguishable from a missing field at the wire level, so an explicit empty variant provides both semantic clarity and allows for runtime validation. With this, the implementation can distinguish between the caller intentionally choosing a scope with no parameters and the caller accidentally forgetting to set the scope parameters. In the context-bearing scope (\texttt{WITH\_CONTEXT} for signatures, all algorithms accept \texttt{SignatureDomainContext} during a signature operation. The \texttt{AsymmetricEncryptParams} variant (carrying an optional \texttt{label} for RSA-OAEP) scopes follow the same structural convention. Because the parameter structure is scope-level, transforming a key from EdDSA to ML-DSA within the context-bearing scope requires no change to the application code. For example, the call \texttt{Sign("signing-key", data, context)} is identical regardless of the underlying algorithm.

The \texttt{Vendor*} variants provide an extensibility point. Provider-specific parameters can be conveyed without modifying the core message definitions. This ensures that the API accommodates proprietary extensions without compromising the type safety of standard operations. The \texttt{user\_context} field (a typed map at field~15) propagates application-defined metadata through the cryptographic operation and into the response. This enables traceability without polluting the cryptographic parameter space.

\subsubsection{Parameter Classification.}

The operation design rests on a three-category classification that assigns each parameter to the party best positioned to handle it correctly:

\begin{enumerate}

\item \textbf{Algorithm parameters} (needed during key creation and associated with the template): key size, curve, hash function, padding etc. These parameters never appear in operation requests.

\item \textbf{System-generated values} (needed per operation and managed by the provider): IVs, nonces, salts, etc. These are generated internally, returned in \texttt{OperationMetadata} during a producer operation such as signing, and are expected to be provided during the corresponding consumer operation such as verification.

\item \textbf{Operational parameters} (needed per operation and provided by the application): AAD, context, tweaks, KDF inputs, etc. They encode application-specific meaning that the system cannot generate on its own and has to be provided by the user/application

\end{enumerate}

This classification addresses a significant class of cryptographic implementation errors. Lazar~et~al.~\cite{lazar2014} found that 83\% of cryptographic vulnerabilities stemmed from API misuse. Our parameter classification reduces the attack surface by ensuring each parameter is handled by the party best positioned to manage it correctly. Type safety is enforced through dedicated message types per scope.

\subsubsection{Operation Metadata.}

Every response includes \texttt{OperationMetadata}, containing the metadata (key version and system-generated values) required to reverse the operation: 

\begin{lstlisting}

message OperationMetadata {
  int32 key_version = 1;
  ProviderOutput provider_output = 2;
  string api_version = 3;
  map<string, string> user_context = 4;
}

message ProviderOutput {
  string encoding = 1; // "raw","der","jws"
  oneof algorithm_output {
    NoAlgorithmOutput no_output = 10;
    AeadOutput aead_output = 11;
    BlockCipherOutput block_cipher = 12;
    CounterModeOutput counter_mode = 13;
    StreamCipherOutput stream_cipher = 14;
    KdfOutput kdf_output = 15;
    VendorOutput vendor_output = 16;
  }
}

\end{lstlisting}

{\emergencystretch=2em

The metadata design reflects security and type-safety choices. The \texttt{ProviderOutput} message carries system-generated values that are produced by the provider during the forward operation and required during the reverse operation. Instead of using a generic map, the design uses a typed \texttt{oneof algorithm\_output} with dedicated messages per algorithm category. For example, \texttt{AeadOutput} contains \texttt{nonce} and \texttt{tag\_length\_bytes}, \texttt{BlockCipherOutput} contains \texttt{iv}, \texttt{CounterModeOutput} carries \texttt{counter\_block} and \texttt{counter\_bits}. This typed approach provides compile-time guarantees that the system-generated values for each algorithm category are correctly structured. The \texttt{VendorOutput} variant (a \texttt{map<string, bytes>}) provides an extensibility option for proprietary providers without compromising type safety for standard algorithms. The \texttt{encoding} field records the output format (\texttt{"raw"}, \texttt{"der"}, \texttt{"jws"}, \texttt{"cbor"}) if required. The \texttt{NoAlgorithmOutput} variant serves as an explicit marker for operations that produce no system-generated parameters (e.g., signatures and MACs), distinguishing the case where no output is produced from output being omitted in error.

Three fields are \textit{intentionally excluded} from the operation metadata: \texttt{template\_id}, \texttt{provider\_id}, and \texttt{timestamp}. Because the metadata is not cryptographically signed, including these fields would create a false sense of auditability. A tampered \texttt{template\_id} could mislead downstream systems into using incorrect decryption parameters. The fields that \textit{are} present in the metadata are self-authenticating. Tampering to provide a wrong \texttt{key\_version} or a wrong initialization vector, for example, causes the AEAD authentication tag check to fail. Tamper detection is provided through the cryptographic primitive itself rather than through a separate integrity mechanism. Audit systems that require \texttt{template\_id} or \texttt{timestamp} should obtain them through the \texttt{ReadKey} RPC or through an independent logging infrastructure, respectively. The \texttt{user\_context} map propagates the application-defined metadata supplied in the request as is, enabling end-to-end traceability.\par}

\subsection{Provider Decoupling (C3)}
\label{sec:api-provider}

The API addresses provider coupling by structurally separating algorithm specific aspects from implementation specific aspects. For example, AES-256-GCM provides 256-bit key security against brute-force search, regardless of the implementation. This is a mathematical property independent of where the computation executes. On the other hand, whether a particular implementation is FIPS~140-3 certified is a deployment fact determined by the provider. This satisfies the principles P8 (Algorithm — Implementation Separation), P9 (Implementation Portability), and P13 (Discoverable Extensibility).

\subsubsection{Two-Registry Architecture.}

The API supports two independent registries. The \textit{template registry} defines immutable and named configurations for algorithms:

\begin{lstlisting}
message TemplateInfo {
  string template_id = 1;
  string display_name = 2;
  string description = 3;
  repeated ScopedCapabilities scoped_capabilities = 4;
  AlgorithmDetails algorithm = 5;
  TemplateStatus status = 6;     // Lifecycle
  string deprecation_notice = 7;
  repeated string standards = 8; // RFCs, FIPS
  CycloneDXAlgorithmProperties cyclonedx = 9;
}
\end{lstlisting}

Each template declares its \texttt{scoped\_capabilities}. These are the scopes it can satisfy, together with the cryptographic operations it supports within each scope and the security guarantees it provides. This structure enables a single template to serve multiple scopes (e.g., an ECDSA template supports both context-bearing and standard signature scopes) and makes the template-to-scopes mapping explicit and queryable. The \texttt{ScopedCapabilities} message bundles a \texttt{ScopeSpecification}, a list of supported \texttt{CryptoOperation} values, and a \texttt{SecurityGuarantees} record (containing formal security notions such as IND-CCA2 and EUF-CMA, practical outcomes, and operational bounds). In this design, security guarantees are \textit{per-scope} rather than per-template. This is because the same algorithm may provide different guarantees in different usage contexts. The \texttt{standards} field lists applicable standard references (e.g., FIPS~197, RFC~5116). The \texttt{CycloneDXAlgorithmProperties} message bridges the API's internal vocabulary with the CycloneDX CBOM interchange format~\cite{cyclonedx}, carrying fields such as \texttt{nist\_quantum\_security\_level}, \texttt{algorithm\_family}, and \texttt{parameter\_set\_identifier} and enables automated CBOM generation directly from template metadata.

The \textit{provider registry}, on the other hand, defines execution backends as dynamic, deployment-specific configurations. It makes a key distinction between a \textit{provider type} (e.g., ``SoftHSM~2.x'') and a \textit{provider instance} (e.g., ``hsm-cluster-01 running SoftHSM~2.x at partition~3'').

\begin{lstlisting}
message ProviderInfo {
  string provider_id = 1;
  string display_name = 2;
  string description = 3;
  ImplementationProperties default_implementation = 10;
  repeated ProviderTemplateSupport template_support = 11;
  ProviderType provider_type = 12;
}

message ProviderTemplateSupport {
  string template_id = 1;
  ImplementationProperties implementation = 2;
  string notes = 3;
  bool enabled = 4;
}
\end{lstlisting}

The template registry captures the static algorithm-specific properties, while the provider registry captures the implementation-specific properties. Each provider carries a \texttt{default\_implementation} record of type \texttt{ImplementationProperties}, which encodes deployment-specific facts across five categories: certifications (FIPS~140 level and certificate number, Common Criteria), code quality (formal verification, verification framework such as F* or Coq, memory-safe language), side-channel resistance (constant-time execution, side-channel hardening), performance (hardware acceleration, specific hardware features such as AES-NI or AVX2), and security posture (known CVEs). The \texttt{ProviderTemplateSupport} field can \textit{override} the provider's defaults on a per-template basis. For example, a provider may be generally FIPS~140-3 Level~1 certified but hold Level~3 certification specifically for its AES-GCM implementation. This override pattern enables precise representation of real-world certification without requiring separate provider definitions for each certification type.

An implementation can join the two registries at runtime. An example implementation for selecting a provider could look like the following: (1)~filter templates by scope; (2)~apply policy constraints; (3)~match providers by template support; (4)~filter by provider requirements; and (5)~rank by match score.

\subsubsection{Constraint-Based Provider Selection.}

Applications can express their need for specific execution requirements through typed constraints and do not have to name providers explicitly. Naming providers is also supported:

\begin{lstlisting}
message MatchProvidersRequest {
  string template_id = 1;
  ProviderRequirements requirements = 2;
  bool enabled_only = 3;
}

message ProviderRequirements {
  optional bool fips_140_certified = 1;
  optional Fips140Level min_fips_level = 2;
  optional bool common_criteria_certified = 3;
  optional bool formally_verified = 4;
  optional bool memory_safe = 5;
  optional bool constant_time = 6;
  optional bool side_channel_hardened = 7;
  optional bool no_known_cve = 8;
  optional bool
    prefer_hardware_accelerated = 9;
  map<string, string> additional = 15;
}
\end{lstlisting}

The \texttt{MatchProviders} RPC can be used to filter by constraints such as using a certified implementation. By specifying the constraint, the application code does not encode a deployment-specific provider. This constraint-based approach is essential for portability. The same application code deployed in a development environment (software provider) and a production environment (HSM provider) requires only different provider configurations, not different application logic.

As mentioned previously, the provider model distinguishes between a \textit{provider type} and a \textit{provider instance}. A provider type (e.g., ``PKCS\#11'' or ``SoftHSM~2.x'') is a static definition that can shipped with the system. It declares the capabilities and a configuration schema for using the provider. A provider instance (e.g., ``hsm-payments-slot5'') is a runtime configuration referencing a provider type with specific deployment parameters ( such as connection endpoint, partition identifiers, etc.). This is similar to the class-instance relationship in object-oriented design and meets typical operational use cases. Organizations routinely deploy multiple instances of the same provider type with different configurations and certification levels. 
The \texttt{ProviderService} accordingly separates provider type queries (\texttt{ListProviders}, \texttt{GetProvider}) from instance management (\texttt{RegisterProviderInstance}, \texttt{ListProviderInstances}, \texttt{UpdateProviderInstance}, \texttt{DeleteProviderInstance}).

\subsection{Decoupling Mechanism (C4): Policy as an Abstract Object (C4)}
\label{sec:api-policy}

The API reaches the C4.3 (policy support for cryptographic governance) level of the assessment framework through a format-agnostic policy abstraction, that realizes the principles P2 (Temporal Decoupling) and P10 (Policy-Engine Independence).

\subsubsection{Policy as Interface, Not Implementation.}

The API defines an interface for policy lifecycle management without mandating a policy language, schema, or implementation requirement, so that organizations can use their own dialect for cryptographic governance. In other words, the API aims at integrating existing governance systems, so that the governance layer is not impeding adoption. Policies are thus referenced by name and their content are raw bytes, interpreted according to the engine's implementation language. Note that the \texttt{policy} field in \texttt{CreateKeyRequest} is a named reference to a policy. Though the API does not specify a policy language, the interface describes functional requirements for a policy engine: it must resolve scopes into templates, allow or deny specific templates, enforce key lifecycle constraints, and define approved providers. An implementation of the policy engine may use any existing format or tools, such as YAML files, OPA/Rego rules, or HashiCorp Sentinel policies.

\begin{lstlisting}
service CryptoPolicyService {
  rpc CreateCryptoPolicy(...)
    returns (...);
  rpc ReadCryptoPolicy(...)
    returns (...);
  rpc UpdateCryptoPolicy(...)
    returns (...);
  rpc DeleteCryptoPolicy(...)
    returns (...);
  rpc ListCryptoPolicies(...)
    returns (...);
  rpc EvaluatePolicy(...)
    returns (...);
  rpc BatchEvaluatePolicy(...)
    returns (...);
}
\end{lstlisting}

The \texttt{EvaluatePolicy} RPC provides a way to programmatically introspect policies and evaluate the outcome of a policy enforcement on different cryptographic operations. The API specifies a policy name, an intended operation (from a typed \texttt{CryptoOperation} enumeration covering all the RPCs), and the relevant context (key name, template identifier, scope specification, or provider identifier). The response returns a boolean \texttt{allowed} verdict, a typed \texttt{PolicyDenialReason} enumeration (with values such as \texttt{TEMPLATE\_NOT\_ALLOWED}, \texttt{SCOPE\_NOT\_ALLOWED}, \texttt{PQC\_REQUIRED}, \texttt{SECURITY\_LEVEL\_INSUFFICIENT}), a human-readable explanation. It also returns the \texttt{PolicyDecisionDetails} containing the list of checks performed, the failed check, and the set of templates and operations that \textit{would} be permitted under the current policy. An implementation of the API can choose the granularity of detail that it wants to convey. The \texttt{BatchEvaluatePolicy} variant evaluates multiple requests in a single call, returning per-request results alongside aggregate counts of allowed and denied operations. For example, this can be used to enable efficient pre-migration impact assessment when an administrator needs to determine how many existing keys would be affected by a policy change.

\subsubsection{Four Governance Responsibilities.}

The policy engine must address four distinct governance concerns, each independently configurable:

\begin{enumerate}
\item \textbf{Scope-to-template mapping.} When an application creates a key with scope \texttt{AEAD\_STANDARD}, the policy resolves this to a concrete template (e.g., \texttt{aes-256-gcm-96}). Changing the mapping, for instance, to Chacha20-Poly1305 requires only a policy update, with no code changes propagating to any application. This is the mechanism that transforms algorithm selection from a development decision into a governance decision.

\item \textbf{Allow/deny list enforcement.} Templates in the \textsc{forbidden} lifecycle state must be blocked regardless of policy, providing defense-in-depth. This two-layer enforcement ensures that even a misconfigured policy cannot select a banned algorithm. This addresses the common concern that policy-driven selection can lead to the selection of weak algorithms.

\item \textbf{Key lifecycle constraints.} Exportability, maximum version count, rotation schedules, and version retention are governed through policy, and not hardcoded. Different organizational units may require different lifecycle parameters for keys serving the same cryptographic purpose.

\item \textbf{Provider constraints.} Provider constraints must be possible to configure. For example, signing keys are required to use HSM-backed providers, but MAC keys may use software providers.

\end{enumerate}

\subsubsection{Two-Phase Evaluation.}

Policy evaluation is done at two distinct phases. Both serve to accomplish different goals of governance. 
1st phase: Upon the creation of keys, the policy engine verifies the request, maps the scope to an appropriate template, and selects a provider. 
2nd phase: When the keys are used to perform a cryptographic operation, the policy engine will evaluate the current policy version, which may be a new version since the key was created. The re-evaluation process is important to ensure timely governance. In other words, if a deprecated algorithm is forbidden by an updated policy, then no additional forward operations (encryption, signatures) can use the previously created key (bound to that deprecated algorithm). However, reverse operations (decrypting, verifying) remain allowed during the transition period so that data remain accessible.

Policy engine’s decision to treat forward and backward operations asymmetrically was made based on the difference between creating new cryptographic commitments versus using existing ones. Organizations can immediately deprecate algorithms from future use; but provide organizations a configurable amount of time to recover their data using the old algorithm. In order for changes in the policy to reflect back onto keys that were created prior to such a policy change and for the policy to act as a source of authority for governance, it is critical that policy evaluations occur every time keys are used.

\subsection{Decoupling Authority (C5)}
\label{sec:api-authority}

The API enables both access authority and cryptographic governance by modeling all endpoints as resources amenable to IAM controls. This realizes P7 (Separation of Concerns).

\subsubsection{Service Architecture.}

The API groups RPCs into 7 gRPC services, each representing a distinct authority domain. For example, the policy service aims at governing cryptography through policy updates and evaluation, the key management service allows to create and transform keys, and the crypto service serves cryptography consumption. Each of these capabilities require different levels of trust. The service decomposition enables coarse-grained access control by granting access to specific services only. 

\begin{lstlisting}
// Single-shot cryptographic operations (12 RPCs)
service CryptoService {
  rpc Encrypt(...) returns (...);
  rpc Sign(...) returns (...);
  rpc GenerateMAC(...) returns (...);
  rpc Digest(...) returns (...);
  rpc GenerateRandom(...) returns (...);
}
// Multi-part and streaming operations (43 RPCs)
service StreamingCryptoService {
  rpc EncryptInit(...) returns (...);
  rpc EncryptUpdate(...) returns (...);
  rpc EncryptFinal(...) returns (...);
  // and other message-based, dual-function patterns
}
// Key lifecycle management (11 RPCs)
service KeyManagementService {
  rpc CreateKey(...) returns (...);
  rpc RotateKey(...) returns (...);
  rpc TransformKey(...) returns (...);
  rpc MigrateKey(...) returns (...);
  rpc ValidateKeyOperation(...) returns (...);
}
// Key establishment operations (6 RPCs)
service KeyEstablishmentService {
  rpc DeriveKey(...) returns (...);
  rpc WrapKey(...) returns (...);
  rpc EncapsulateKey(...) returns (...);
}
// Policy lifecycle and evaluation (7 RPCs)
service CryptoPolicyService {
  rpc EvaluatePolicy(...) returns (...);
  rpc BatchEvaluatePolicy(...) returns (...);
}
// Algorithm catalog and scope discovery (3 RPCs)
service AlgorithmDiscoveryService {
  rpc ListTemplates(...) returns (...);
  rpc GetTemplate(...) returns (...);
  rpc ListScopes(...) returns (...);
}
// Provider catalog and matching (8 RPCs)
service ProviderService {
  rpc MatchProviders(...) returns (...);
  rpc RegisterProviderInstance(...) returns(...);
}
\end{lstlisting}

{\emergencystretch=1em
The seven-service decomposition is significant for three reasons. First, services with different trust requirements are physically separated: an identity granted access to \texttt{CryptoService} (routine operations) cannot invoke \texttt{TransformKey} on \texttt{KeyManagementService} (a privileged key mutation) or \texttt{RegisterProviderInstance} on \texttt{ProviderService} (infrastructure configuration), because those RPCs reside in distinct service definitions with independent authorization policies. Second, the separation of \texttt{CryptoService} (single-shot RPCs) from \texttt{StreamingCryptoService} (multi-part RPCs) reflects an operational distinction: streaming operations require server-side session state and present different resource-exhaustion risks, warranting independent rate limiting and monitoring. Third, read-only discovery services (\texttt{AlgorithmDiscoveryService}, \texttt{ProviderService}) can be exposed to broader audiences, including automation tools and CBOM generators, without granting access to cryptographic operations or key mutations.\par}

\subsubsection{Role-Based Separation.}

Some roles may need to access only subsets of services. For example, \textit{Users} access \texttt{CryptoService}, \texttt{StreamingCryptoService}, and \texttt{KeyEstablishmentService} to compute cryptographic operations, but only the \texttt{ReadKey} RPC of \texttt{KeyManagementService} to obtain keys' metadata, while \textit{Admins} access \texttt{KeyManagementService} for privileged key mutations (transform, migrate, destroy), \texttt{CryptoPolicyService} for governance, and \texttt{ProviderService} for infrastructure configuration. In this specific example, both \texttt{Users} and \texttt{Admins} can access the \texttt{KeyManagementService}, but \texttt{Users} must be restricted to the \texttt{ReadKey} RPC. Systems implementing the API are responsible for defining proper access control. They can leverage service-based access control when roles can access an entire service, define access control per endpoint or RPC when roles can access only subsets of services, or use resource-based access control when roles can access only specific resources within a service or endpoint.

\subsubsection{Temporal Separation.}

Three decision layers operate independently:

\begin{enumerate}
\item \textbf{Code authorship time.} Developers write \texttt{Sign(key\_name, input)} with no algorithm or implementation knowledge baked into that request. The operation interface remains stable across changes of algorithms and/or providers (satisfying the P1 principle).

\item \textbf{Deployment time.} Administrators create keys with scope-based specifications and configure policies.


\item \textbf{Governance time.} Security teams modify policies and convert/modify key(s), which may include changing an algorithm without requiring any modification to the existing application code

\end{enumerate}

\subsection{Agility Enablers: Key Evolution and Provider Migration (E1, E2)}
\label{sec:api-enablers}

In the companion paper, no evaluated cryptographic API provides E1.3 (algorithm transformation) via an explicit operation that can be independently authorized. The API addresses this through three distinct evolution operations. This realizes the principles P5 (Key Abstraction) and P6 (Key Evolution).

\subsubsection{Three Evolution Operations.}

\begin{lstlisting}
// Same algorithm, new material (key hygiene)
message RotateKeyRequest {
  string name = 1;
}

// New algorithm, same identity (agility)
message TransformKeyRequest {
  string name = 1;
  bool retain_key_bytes = 2;
  ScopeSpecification scope_spec = 3;
  string template_id = 4;
}

// New provider, same identity (migration)
message MigrateKeyRequest {
  string name = 1;
  oneof target {
    string target_instance_id = 2;
    ProviderTarget provider_target = 3;
  }
  MigrationStrategy strategy = 4;
  // Optional: also change alg
  ScopeSpecification scope_spec = 5;
  string template_id = 6;   
}

enum MigrationStrategy {
  UNSPECIFIED = 0;
  PROVIDER_SWITCH = 1;     // Metadata-only
  EXTRACT_AND_IMPORT = 2;  // Export+import
  WRAPPED_TRANSFER = 3;    // Wrapped transit
  REKEY_AND_ARCHIVE = 4;   // Rekey, keep old
  REKEY_AND_DESTROY = 5;   // Rekey, destroy
}
\end{lstlisting}

\textbf{Rotation} (\texttt{RotateKey}) generates new key material under the same algorithm. Rotation must create a new version while preserving previous versions. The request is minimal and requires only a key name. This is because all algorithm parameters required for rotation are inherited from the existing key. Rotation is a routine operation that addresses key hygiene and exposure limits.

\textbf{Transformation} (\texttt{TransformKey}) is the operation that enables algorithm agility, required for post-quantum transition. A transformation generates a new key version with a different algorithm or with the same algorithm but different parameters. The API supports either a scope specification that policy resolves to a concrete template, or an explicit template identifier for the transformation operation. The \texttt{retain\_key\_bytes} flag enables changing the algorithm or its configuration while keeping the same key bytes. Such scenario may happen when the new template is mathematically compatible with the existing key material (e.g., changing only the hash algorithm used with a signature scheme). An example of transformation includes transforming a key from ECDSA P-256 to ML-DSA-65, which creates a new version while preserving the key name and ECDSA version history. The application code that calls \texttt{Sign("signing-key", data)} remains unchanged and independent of the underlying algorithm used by the "signing-key".

\textbf{Migration} (\texttt{MigrateKey}) moves a key between providers. The \texttt{oneof target} accepts either a specific provider instance identifier or a \texttt{ProviderTarget} with typed requirements, enabling both precise targeting and constraint-based selection. Five strategies meet the security constraints of different provider types: \textsc{provider\_switch} performs a metadata-only update for providers that share a keystore; \textsc{wrapped\_transfer} protects key material during transit; \textsc{rekey\_and\_archive} and \textsc{rekey\_and\_destroy} address HSM non-exportable keys by generating new material at the target.

The optional \texttt{key\_specification} field (not shown in the listing) in \texttt{MigrateKey} enables simultaneous algorithm and provider change. It allows migrating to an HSM while upgrading from ECDSA to ML-DSA in a single atomic operation, avoiding an intermediate state in which the key exists on the old provider with the new algorithm, or vice versa.

Because transformation and migration are high-stakes operations that may be irreversible (particularly when the source material is destroyed), the API provides a \texttt{ValidateKeyOperation} RPC for pre-flight dry-run validation, which an implementation may choose to support. Given a key name and an intended transform or migrate operation, it returns the current key state, a list of feasible strategies with security annotations, and a recommended strategy with justification. Each \texttt{StrategyOption} reports whether the strategy is feasible, the reason for any infeasibility (e.g., ``source provider does not support key extraction''), security notes (e.g., ``key material will transit network in wrapped form''), and a complexity rating. This enables administrators to assess the impact and feasibility of evolution operations before committing to them. This can be particularly valuable during large-scale post-quantum migration campaigns, where thousands of keys may need to be transformed.

\subsubsection{Version History and Data Continuity.}

All three operations preserve version history, ensuring data encrypted or signed under previous versions remains accessible. Each key version is expected to maintain its algorithm association, provider binding, and lifecycle state independently. The \texttt{OperationMetadata} produced by cryptographic operations records the key version used, enabling correct decryption or verification under any historical version. As an example, consider a signing private key that is initially associated with ECDSA-P256. When the API produces a signature with the key, it includes 1 as version number for the key in the operation metadata. Later, if the key is transformed into a ML-DSA-65 key, new signatures will include 2 as version number in the metadata. Upon verification, the API uses the metadata to retrieve the key version.

Keys follow six lifecycle states aligned with NIST SP~800-57~\cite{nist-sp800-57}: \textsc{pre-active}, \textsc{active}, \textsc{deactivated}, \textsc{compromised}, \textsc{destroyed}, and \textsc{archived}. Forward or producer operations (encrypt, sign, etc.) are limited to active versions of keys, while reverse or consumer operations (decrypt, verify, etc) may be performed on both active and deactivated key versions. This allows for graceful migration from one algorithm to another by deactivating the old version after a transformation, that can no longer sign new data but still validate existing signatures.

\subsection{Discovery: Runtime Catalog and Pre-Flight Validation}
\label{sec:api-discovery}

The framework requires that new algorithms are supported without API changes. The API satisfies this through discovery services and realizes the principles P12 (Property Taxonomy) and P13 (Discoverable Extensibility).

\subsubsection{Algorithm Discovery Service.}

\begin{lstlisting}
service AlgorithmDiscoveryService {
  rpc ListTemplates(ListTemplatesRequest)
    returns (ListTemplatesResponse);
  rpc GetTemplate(GetTemplateRequest)
    returns (GetTemplateResponse);
  rpc ListScopes(ListScopesRequest)
    returns (ListScopesResponse);
}
\end{lstlisting}

\texttt{ListTemplates} uses the exact same \texttt{ScopeSpecification} as is done with \texttt{CreateKeyRequest}. That Scope Specification serves as a filter on the set of templates being listed. Thus, it allows for \textit{discovery-before-creation} of keys. Before requesting the creation of a new key, applications or workflow automation tools can discover which templates are available to them based on their desired scope and security requirements. The \texttt{ListTemplates} request also takes two additional parameters: \texttt{allowed\_statuses}, which can be used to select only those templates whose lifecycle status matches one of the provided values, and \texttt{required\_standards}, which can be used to select only those templates that meet at least one of the specified standard compliance levels (for example: "FIPS~204"). These capabilities enable compliance-aware filtering when enumerating templates. Because \texttt{ScopeSpecification} is reused throughout the creation, discovery, and transformation operations, it provides a common language that governs all three operations. This prevents divergence in terms of how templates can be enumerated versus requested. \texttt{ListScopes} returns the full scope taxonomy and includes descriptive information about each scope along with a typical use case for that scope. As such, \texttt{ListScopes} enables self-documenting APIs and automatically generated workflows.

\subsubsection{Provider Discovery.}

{
\emergencystretch=3em
Provider matching uses the \texttt{MatchProviders} RPC (Section~\ref{sec:api-provider}) with the same typed \texttt{ProviderRequirements} constraints can be used to discover registered providers.\par}

\subsubsection{Discovery as an Agility Enabler.}

Adding a new post-quantum algorithm to an implementation of the API requires four steps: (1)~add a template to the static catalog; (2)~verify the addition via \texttt{ListTemplates}; (3)~update the policy mapping to support the newly added algorithm; (4)~register or update the provider code to implement the new algorithm and (5)~optionally transform existing keys. The application code that consumes this API need not be changed. The abstract API enables this agility, and the discovery APIs provide the observability layer for verification.

\subsection{Deployments}

The API does not restrict the deployment mode of its implementation. It could be deployed in five different ways, including as a library that is completely independent of networks, to  a fully centralized service. The important invariant is that {\em application code remains unchanged across all five deployment options}. Changing the way you deploy your API is therefore a matter of configuring infrastructure, not altering code.

{\emergencystretch=2em

\noindent\textbf{Mode~1 (Standalone)} executes the full stack --- i.e., cryptographic execution, key storage, and policy evaluation --- entirely inside the application process. As a result, in this mode, the policy is loaded into the application from local configuration files. Thus, all security logic and cryptographic control remain local to the application. Because there are no network dependencies, the latency associated with cryptographic operations will be minimal. Therefore, Mode~1 would be well-suited for embedded systems, air-gapped environments, and other settings where Internet connectivity is either severely restricted or simply undesirable.

\textbf{Mode~2 (Policy Managed)} includes remote policy governance but keeps both key material and cryptographic execution local to the application. In this case, the central authority pushes/pulls/hybrid synchronizes policy updates and governance decisions. Nonetheless, actual cryptographic operations and key management are still executed locally. Thus organizations can centrally manage and update policies (for example, allowed algorithms or usage restriction policies) without centralizing key custody. This balances governance needs with local control and security.

\textbf{Mode~3 (Lifecycle Managed)} extends centralization further by incorporating remote lifecycle orchestration. In this case, a central server has the authority to command only lifecycle events such as key rotation, deactivation, or transition to a new algorithm. The keys themselves will continue to reside in local storage. Similarly, the cryptographic operations will continue to be performed within the application. The major difference in Mode~3 is that we distinguish between orchestration (i.e., Key lifecycle management, policy management) and custody (i.e., holding the key material). Only orchestration is centralized, not custody or execution.

\textbf{Mode~4 (Key Managed)} centralizes key custody. In this mode key material is also stored and managed by a central server. When the application needs to execute a local cryptographic operation, it must now transport keys securely. To accomplish this a strong mechanism such as TLS mutual authentication must be used for secure communication between the application and server. Additionally, TEEs may be used for remote attestation and releasing keys closer to the application. By removing key custody from the application host, Mode~4 allows stronger separation of duty and enables better compliance with regulations regarding key management.

\textbf{Mode~5 (Fully Centralized)} is the most centralized deployment. In this case all cryptographic operations, key storage, policy enforcement are performed by a remote service. The application behaves as a thin client and delegates all cryptographic function calls to a central service. This deployment provides the best auditable record and enables centralized governance over all cryptography activities. The price paid for this deployment model is higher latency due to each cryptographic operation requiring one rounds trip to/from the remote service. For performance sensitive applications this increased latency may have negative impacts on system performance.

The five deployments create a spectrum of increasing levels of centralization and organizations can use the right centralization required by their compliance posture and performance requirement. The ability to move up/down the spectrum of increasing/decreasing levels of centralization without modifying application code directly results from the abstraction the API provides. Regardless of whether or not the implementation resides in a local environment versus a remote environment the application always communicates with the same abstract interface.

\section{Evaluation}
\label{sec:evaluation}

In this section, we evaluate the API against the assessment framework of the companion paper ~\cite{paper1-framework}. For each dimension, we
state the level reached and the design elements that justify it. The levels
reached are C1.3, C2.2, C3.3, C4.3, C5.3--5.4, E1.3, and E2.2.

\subsection{Operation coupling (C1).}

The API reaches level~C1.3
(scope-consistent operations). Every operation exposes a uniform function signature, keys abstract algorithm identity, and the API handles system-generatable parameters. Operational parameters remain part of the request's parameters but are consistent across scopes: if a value of the \textit{scope\_params} field of the request is accepted by an algorithm of a given scope, then it should be accepted by any algorithm of that scope. The level C1.4 (fully managed operation) is not reached as the API does not enable primitive-level agility (ie. substitution of algorithms sharing the same primitive). This is a tradeoff between agility and flexibility, with scopes defining subsets of primitives with consistent operational parameters to increase flexibility.  

\subsection{Key creation coupling (C2)}

The API reaches level~C2.2
(intent-based creation). At its highest abstraction level, an application supplies
only a scope that describes its cryptographic intent, declaring what the key must achieve
without naming an algorithm and its configuration (Section~\ref{sec:binding-chain} and~\ref{sec:api-creation}).
This intent can be strengthen with universal
security properties (e.g., \texttt{quantum\_safe} \texttt{fips\_approved}) and scope-specific properties (e.g., deterministic for signatures), narrowing the set of acceptable
templates while still leaving the final algorithm choice to the system. A fallback option to name an explicit template allows to select a
concrete algorithm and its configuration when fine-grain control is needed. 
The design does not reach C2.3 (purpose-driven requests), because the intent is
expressed in cryptographic rather than business terms. 

\subsection{Provider coupling (C3)}

The API reaches level~C3.3 (key-driven
provider selection). Providers are bound to the keys at creation time,
and application code remains provider-agnostic: operation requests are identical
for keys bound to different providers (Section~\ref{sec:api-provider}). Upon key creation, the provider can be selected explicitly by specifying its identifier or constrained to match requirements
(e.g., FIPS~140-3 level, constant-time execution). The verdict, among providers matching the requirements, is given by the policy rules, that may enforce additional provider requirements. At operation, the provider resolution is abstracted behind the key, and different keys may be bound to different providers, which are the requirements of C3.3. However, the design rationale enforces one provider per key version, meaning that the per-operation provider routing requirement of C3.4 (full provider agnosticism) cannot be satisfied.

\subsection{Decoupling mechanism (C4)}
The API reaches level~C4.3 (policy
engine). Because the key abstracts both the algorithm and the provider when cryptographic operations are computed, the coupling is entirely determined by the key. It means that all the couplings---both the template and provider---are determined at key creation. The decoupling mechanism must resolve two abstractions: scopes into concrete templates and provider requirements into concrete provider instances (Section~\ref{sec:api-policy}). Both are resolved by the policy engine. The functional interface for policies mandates supporting expressive constraints for templates, providers and key lifecycle management, thus meeting the requirement of C4.3. At the highest abstraction level, a key creation request only specifies a scope, that is resolved into a pair of template and provider by the policy engine. The only binding that remains between the cryptography and the application is the cryptographic intent.

\subsection{Decoupling authority (C5)}
The level reached depends on the concrete policy language and range from 5.3 (scoped governance
domains) to 5.4 (federated governance). At a minimum, policies
must define compliance requirements in terms of templates and
providers. Services for cryptographic operations, key creation operations and policy management operations are separated to allow for service-level access control. It is then straightforward to define role-based access control (level 5.2), where security officers access the policy service to control the set of approved algorithms, while developers access the crypto service to compute cryptographic operations. C5.3 is achieved because the policy engine supports many independent policies, that are referenced by name. Each governance domain can thus define and maintain its own policy, independently of other domains. When policies can hierarchically reference one another, enabling policy inheritance, the dimension reaches level 5.4 (federated governance). 
\\ 

\subsection{Agility enablers (E1, E2)}

The API exposes dedicated, independently
authorizable RPCs for key evolution (Section~\ref{sec:api-enablers}).
\texttt{TransformKey} is a first-class operation that generates new material under
a different target algorithm while preserving the key identifier and version
history, achieving E1.3 (explicit algorithm transformation). \texttt{MigrateKey} orchestrates the
transfer of a key between providers as a single coordinated operation with
identifier preservation and a choice of transfer strategies, achieving E2.2
(explicit migration API). Additionally, the pre-flight \texttt{ValidateKeyOperation} RPC further
allows the feasibility and security implications of such operations to be assessed
before they are committed. The API design does not specify policy-driven algorithm transformation (E1.3) nor provider migration (E1.3): automated triggering of transformation and migration is the responsibility of the system relying on the API.

\subsection{Summary}

\textbf{Summary.} Across the seven dimensions, the intent-based API attains
C1.3, C2.2, C3.3, C4.3, C5.3--5.4, E1.3, and E2.2. These levels close the
three gaps found by the companion paper among deployed
systems~\cite{paper1-framework} (intent-based key creation (C2.2), cryptographic
governance distinct from access control (C5.3+), and explicit algorithm
transformation (E1.3)). The choice to stop short of the maximal level on a dimension (C1.3 rather than C1.4, C2.2 rather than C2.3, C3.3 rather
than C3.4, and E1.3/E2.2 rather than their policy-driven counterparts) reflects the framework's tradeoff between agility and flexibility. Higher autonomy, coming at the cost of flexibility, can be obtained from an orchestration layer built above the API. Taken together, the levels achieved are sufficient to reduce the post-quantum
transition from a distributed software-engineering effort to an operational
procedure driven by policy updates and controlled key transformations.

\section{End-to-End Migration Scenario: ECDSA to ML-DSA}
\label{sec:migration}

In this section, we walk through a concrete post-quantum migration example to demonstrate how the API enables substituting algorithms without changing the application code. In this example, an organization deploys a distributed application to sign audit log entries. The application code for signature is as follows: 

\begin{lstlisting}
resp := client.Sign(SignRequest{
    key_name: "audit-log-signing-key",
    input:    auditEntry,
    context: []byte("audit-log-v2"),
})
\end{lstlisting}

\noindent\textbf{Step~1: Initial deployment.} Let's assume the key \texttt{"audit-log-signing-key"} is created with a signature scope specification of \texttt{WITH\_CONTEXT}, and the policy maps this scope to Ed25519ctx (a context-bearing signature algorithm). Let's assume the key resides on a software provider.

\noindent\textbf{Step~2: Policy update.} The security team updates the policy to map the \texttt{WITH\_CONTEXT} signature scope to ML-DSA-65. Both Ed25519ctx and ML-DSA-65 accept the same operational parameters (message and context), so they reside in the same scope. New keys created after this point will automatically use ML-DSA-65. No application code changes are required. 

\noindent\textbf{Step~3: Key transformation.} An administrator transforms the existing key:
\begin{lstlisting}
client.TransformKey(TransformKeyRequest{
    name: "audit-log-signing-key",
    scope_spec: ScopeSpecification{
        signature: SignatureScopeSpec{scope: WITH_CONTEXT}    
    template_id: "ML-DSA-65",
    }
})
\end{lstlisting}
A new key version (v2) is created for ML-DSA-65, while the Ed25519ctx version (v1) is retained for verification. New signing operations automatically use v2, without application code changes. 

\noindent\textbf{Step~4: Template deprecation.} The Ed25519ctx template is marked \textsc{forbidden}. No new keys can be created with Ed25519ctx. Existing Ed25519ctx versions remain available for verification only. \textit{Framework: P13 (lifecycle management).}

\noindent\textbf{Step~5: Provider migration.} The organization moves signing keys to an HSM:
\begin{lstlisting}
client.MigrateKey(MigrateKeyRequest{
  name: "audit-log-signing-key",
  target_instance_id: "hsm",
  strategy: REKEY_AND_ARCHIVE,
})
\end{lstlisting}

\noindent Throughout this five-step migration, the application code remained unchanged. The migration was executed entirely through operational procedures alone. A policy update (Step~2), a key transformation (Step~3), a template lifecycle change (Step~4), and a provider migration (Step~5), each of which were independently executed. The temporal separation between code-authorship decisions and governance decisions is what makes this possible.

\section{Related Work}
\label{sec:related}

\subsection{Application-Level Agility in APIs}

Tink~\cite{tink2018} exposes stable primitive-level interfaces (e.g. \texttt{AEAD}, \texttt{digital signature}) that group algorithms by cryptographic function and enable key rotation within a primitive, with versioned keysets that preserve the ability to decrypt or verify data produced under superseded keys. This model achieves the highest operation decoupling among deployed systems, but its primitive-level abstraction excludes algorithms whose operational parameters do not fit the primitive contract---for example, signature schemes requiring a context string (such as Ed25519ctx or ML-DSA with context) and AES-CCM, which is not exposed under the \texttt{AEAD} primitive. In addition, Tink lacks provider agility, as algorithms' implementations are fixed per language, and externalized algorithm selection is not supported. 

SecAlgo~\cite{secalgo} provides stable interfaces covering algorithms that share the same primitive, together with configuration-file-driven selection of both the algorithm and the underlying provider library. Compared to Tink, SecAlgo introduces the provider selectability that Tink lacks, but does not offer a key abstraction that would enable continued use of existing keys across algorithm changes.

EverCrypt~\cite{evercrypt} is a high-performance, cross-platform, and formally verified cryptographic provider that automatically multiplexes operations between the HACL* and ValeCrypt implementations to obtain the best performance available on the target hardware. Its interface is primitive-level and does not yet cover the full surface required by modern applications---for instance, it lacks a common interface for elliptic-curve cryptography. Its application-level agility is comparable to SecAlgo, with the important addition of formally verified implementations.

eUCRITE~\cite{eucrite} exposes common interfaces for signature and encryption and abstracts algorithm choice through three predefined security levels (\texttt{LOW}, \texttt{MEDIUM}, \texttt{HIGH}), each bound to an opinionated default algorithm. The interface was primarily designed for ease of use rather than agility: the mapping from a security level to a concrete algorithm is encoded in source code, so changing the default algorithm still requires modifying and redeploying the library.

None of these systems provides the combination required to change algorithms without touching application code: intent-based key creation for short-lived keys combined with policy-driven algorithm selection, and algorithm transformation for long-lived keys. Our principles and API patterns address these three gaps.

\subsection{Systems-Level Agility}

ELCA~\cite{elca} formulates additional requirements for achieving cryptographic agility at the enterprise level: orchestrated migration across infrastructure units, cryptography configuration driven by centralized policies, and monitoring and auditing capabilities. Cryptographic operations are decoupled from the application and offloaded to a cryptographic service, centrally controlled by policies that determine which algorithm and implementation to use. However, ELCA specifies an API only for TLS and SSL communication and does not cover cryptographic primitives. Our API patterns address this gap by defining concrete requirements and a corresponding interface for primitive-level operations.

Cho~et~al.~\cite{software-defined-crypto} propose a design feature for crypto-agility targeted at cloud-native applications composed of microservices. They introduce software-defined cryptographic policies---listing vulnerable algorithms and mandating specific cryptographic modules or algorithms---that are automatically enforced through CI/CD pipelines, and define cryptographic providers as an abstraction for cryptography execution.

Our work complements these systems-level approaches. ELCA and software-defined cryptography address the orchestration layer while our patterns address the API and governance layer. A complete agility stack requires both: API patterns that support intent-based creation and key evolution (our contribution), combined with orchestration layers for policy distribution and migration coordination.

\subsection{Cryptographic API Usability}

Green and Smith~\cite{green-smith-usability} propose developer-centric heuristics to minimize cryptographic misuse: APIs should be easy to use, easy to learn, and safe by default. Lazar~et~al.~\cite{lazar2014} show that 83\% of cryptographic vulnerabilities in their study stemmed from API misuse rather than flaws in the underlying primitives. Acar~et~al.~\cite{acar2017} reinforce that security requires simplicity of APIs and further demonstrate that feature completeness (e.g., secure key storage, password-based key generation) and documentation are equally decisive for secure outcomes. Our contribution applies these lessons to agility through separation of concerns (Section~\ref{sec:api-authority}) and parameter classification (Section~\ref{sec:api-operation}). The API exposed to developer is kept simple and easy to learn: interfaces are consistent within scopes, keys abstract cryptographic details, and security is delegated to cryptographic policies. 

\section{Conclusion}
\label{sec:conclusion}

\subsection{Summary of Contributions}

This paper derived thirteen API design principles from five foundational architectural properties. Ten of them constitutes the core principles that are necessary to enable cryptographic agility, while the remaining three are supportive principles that improve the system quality. The API introduced in this paper, based on gRPC and Protocol Buffers, realizes all of these principles by introducing five key design elements:

\begin{enumerate}
\item \textbf{A scope vocabulary} to express cryptographic intent rather than specific cryptographic algorithms and configurations. It enables the decoupling from algorithm identity at key creation (addressing the C2.2 gap). Scope boundaries are defined by operational parameter consistency (P4), ensuring that algorithm substitution within a scope cannot cause runtime failures.

\item \textbf{A three-category parameter classification} that assigns algorithm parameters to templates, system-generated values to providers, and operational parameters to applications, reducing the surface area for parameter misuse.

\item \textbf{A two-registry architecture} that separates algorithm specifications (templates) from provider configurations, enabling independent management of algorithms and implementations.

\item \textbf{A format-agnostic policy abstraction} that enables cryptographic governance without prescribing the policy engine (addressing the C5.3 gap), with four governance responsibilities and two-phase evaluation.

\item \textbf{Three-stage key evolution} (rotation, transformation, migration) with version history preservation (addressing the E1.3 gap), enabling cross-algorithm migration without breaking existing data.
\end{enumerate}

An end-to-end migration scenario demonstrated that these patterns enable a post-quantum transition from a distributed software development project to an operational procedure that requires no application code changes.

\end{document}